\documentclass[superscriptaddress,prl,twocolumn,showpacs,floatfix]{revtex4-1}
\usepackage[T1]{fontenc} 
\usepackage[cp1250]{inputenc}
\usepackage{graphicx} 
\usepackage[english]{babel} 
\usepackage{psfrag}
\usepackage[normalem]{ulem}
\usepackage{amssymb, amsmath, amsfonts}
\usepackage{bm}
\usepackage{multirow}
\usepackage{graphicx}
\usepackage{units}
\include{subfigure}
\usepackage[normalem]{ulem} 
\usepackage[usenames,dvipsnames]{color}

\usepackage{subfigure}

{\begin{list}{}%
         {\setlength{\leftmargin}{#1}}%
         \item[]%
}
{\end{list}}

\begin{document}

\title{Extreme value statistics of mutation accumulation in renewing cell populations}

\date{\today}

\author{Philip Greulich}
\affiliation{Mathematical Sciences, University of Southampton, Southampton, UK}
\affiliation{Insititute for Life Sciences, University of Southampton, Southampton, UK}
\author{Benjamin D. Simons}
\affiliation{Cavendish Laboratory, University of Cambridge, Cambridge, UK}
\affiliation{Wellcome Trust-CRUK Gurdon Institute, University of Cambridge, Cambridge, UK}
\affiliation{Cambridge Stem Cell Institute, University of Cambridge, Cambridge, UK}

\noindent
\begin{abstract}
The emergence of a predominant phenotype within a cell population is often triggered by a rare accumulation of DNA mutations in a single cell. For example, tumors may be initiated by a single cell in which multiple mutations cooperate to bypass a cell's defense mechanisms. The risk of such an event is thus determined by the extremal accumulation of mutations across tissue cells. To address this risk, we study the statistics of the maximum mutation numbers in a generic, but tested, model of a renewing cell population. By drawing an analogy between the genealogy of a cell population and the theory of branching random walks, we obtain analytical estimates for the probability of exceeding a threshold number of mutations and determine how the statistical distribution of maximum mutation numbers scales with age and cell population size.


\end{abstract}
\pacs{x} 
\maketitle

\maketitle 

Over the lifetime of an organism, its constituent cells continuously accumulate DNA mutations, which can affect the pathways that control cell proliferation and survival. Yet, due to gene multiplicity or functional redundancy \cite{Knudson1971,genetic_interactions_cancer_cell2011,gen_buffering_hartwell_science,gen_robustness_devisser2003,gen_robustness_Gu_Nature2003}, disruptions of such pathways may often be tolerated within a homeostatic tissue cell population. Evidence from studies of the cancer genome \cite{genetic_interactions_cancer_cell2011,simons_pnas_2016,martincorena_jones_science2016} suggest that the accumulation of a critical number of individually ``neutral'' or ``near-neutral'' mutations may, in many cases, be necessary to trigger a selective survival advantage on cycling cells -- a process called ``genetic'' or ``epistatic buffering'' \cite{genetic_interactions_cancer_cell2011,gen_buffering_hartwell_science,gen_robustness_devisser2003,Moore2005,epistatic_buffering_jasnos_nature2007,Komarova2014}. The resulting proliferative advantage of mutated cells confers clonal dominance \cite{Wagstaff2013,Alcolea2014} which, if sustained long-term \cite{greco_nature_2017}, constitutes a potential tumour-initiating event. 
Crucially, since one cell within a tissue cell population is sufficient to trigger such an event \cite{development_cancer_book}, the risk of this occurring is naturally dominated by the statistics of rare events -- in this case the extreme accumulation of a multiplicity of mutations within a cell, rather than by the cell population averages reaching some level of mutational burden. The statistics of extreme mutation accumulation represents, therefore, a question of both academic and practical interest. 

The normal maintenance of adult renewing tissue, such as the skin epidermis or the gut epithelium, relies on the activity of stem cells, which divide to replenish functional differentiated cells lost through exhaustion or cell death \cite{Alonso2003,Barker_Clevers2007}. In addition to asymmetric divisions, which leave the stem cell population unchanged \cite{potten_1974}, in most of these tissues the frequent, stochastic loss of stem cells is compensated be replacement via neighbors that divide symmetrically \cite{simons_stemcells_nature,Lopez-Garcia2010}. It is on this background, that these long-lived cells acquire mutations that may lead, in turn, to a selective growth advantage.


Historically, efforts to model how the serial acquisition of mutations can drive tumor progression have focused predominantly on population means or have neglected the potential for epistatic buffering \cite{armitage_doll_1954,armitage_doll_1957,tumour_incidence_rates_1995,bozic_antal_nowak_2010}. The impact of stochastic cell fate dynamics on the statistics of rare mutational signatures has remained under-explored. However, recently, numerical studies have shown how maintenance mechanisms reliant on stochastic stem cell self-renewal can protect cell populations from extreme mutational acquisition events \cite{lander_mut-accumul}. These findings have been reinforced by analytical studies based on a specific model of tumor-initiation involving a ``double-hit'' \cite{Shahriyari2013}. However, the statistical basis of cancer risk on rare event phenomena in renewing tissues remains poorly defined. Here, we present a generic theory for how properties of the extreme mutation number distribution scale with age and cell number, and how this determines the risk of accumulating a critical number of mutations.  In particular, we elucidate how drift dynamics of the renewing cell population moderates the strength of fluctuations, diminishing the frequency of rare events. Besides its relevance for assessing the risk of tumor initiation, this theory also generically elucidates how a predominant phenotype can emerge in a cell population (e.g. bacteria) via the epistatic cooperation of individually (near-)neutral mutations.


To model the long-term accumulation of mutations in a renewing cell population we consider a stochastic model closely related to the Moran process in population genetics \cite{Moran1958}. In this model, cells replicate through division, acquire mutations and are lost stochastically while the total number of cells $N$ is maintained constant (the condition of homeostasis). Therefore, we assign a fixed number of 'sites' $i=1,...,N$ to the cells, where a cell at site $i$ is characterized by mutation number, $m_i$. When a cell at site $i$ is lost, at rate $\lambda$, another cell at a random site $j$ simultaneously divides symmetrically, producing a copy with the same mutational signature, to replace the lost cell on site $i$. 
In addition, any cell $i$ with $m_i$ mutations can acquire stochastically an additional mutation at a constant rate $\mu$. Note that in stem cell populations, asymmetric cell divisions, where one of the daughter cells commits to terminal differentiation and loss, leave the configuration of mutations across cells invariant, and so need not to be considered explicitly. Their potential to effect additional mutations through division is incorporated as the mutation rates are decoupled from the loss/replacement rate. For simplicity, we do not distinguish between the loci of mutations in the genome, an approximation that is valid for low net mutational burden. Furthermore, we consider the scenario before transformation into a hyper-proliferative state, in which the mutations' effect on proliferation is neutral. The model dynamics can be written as the process
\begin{eqnarray}
\label{mutaccumul_model_1}
m_i &\xrightarrow{\displaystyle{\lambda}}& m_j \,\, ,\ \\ \label{mutaccumul_model_2} 
m_i &\xrightarrow{\displaystyle{\mu}}& m_i +1 \,\, ,
\end{eqnarray}
where sites $i$ and $j$ are chosen randomly. 

In the following, we will address the risk $\bar P_{N}(m_{c},T)$ that at least one cell in a population of $N$ cells acquires a critical number of mutations $m_c$ after time $t=T$. This corresponds to the probability that the maximal mutation number across the population, $m^*:=\max(m_1,...,m_N)$, reaches $m_c$, which is related to the cumulative distribution function (CDF) of $m^{*}$, $P^*_N(m_c,T) := {\rm Prob}(m^*(T) \leq m_c) = 1-\bar P_N(m_{c},T)$. In particular, we will study the dependence of the CDF's median and mean on $N$ and $T$.


Before addressing the dynamics of mutation accumulation on the background of stochastic cell loss/replacement, as a benchmark, we first consider the case $\lambda=0$ in which cells accumulate mutations independently. In this case, the model describes $N$ independently distributed Poisson processes and an expression for the extreme mutation number distribution can be determined straightforwardly. Although this expression does not admit a simple scaling form \cite{extr-val_leadbetter_book}, the dependence on $N$ can be well-approximated for large $\mu T$ by normally distributed random variables with a mean and variance $\mu T$ (see Supplemental Material). From this correspondence it follows that, at large $N$  and $\mu T$, the difference between maximum mutation number $m^{*}$ and the population mean, $\Delta m^* := m^{*} - \langle m_{i} \rangle$ has a CDF $P^{*}_{N}(\Delta m_{c}) = {\rm Prob}(\Delta m^{*} < \Delta m_{c})$ which follows a Gumbel distribution,
\begin{align}
\label{gumbel_eq}
P^{*}_{N}(\Delta m_{c}) = e^{-e^{-X}} \mbox{with } X=\frac{\Delta m_{c} - \tilde m}{\sigma_{N}} - \ln\ln\,2 \,\,\, ,
\end{align}
with median $\tilde m$ and scaling width $\sigma_{N}$ given by \cite{extr_val_bouchaud}
\begin{align}
\tilde m \approx \sqrt{2 \mu T \ln N}, \hspace{5mm} \sigma_{N} \approx \sqrt{\frac{\mu T}{2 \ln N}} \,\,\, .
\end{align}
The scaling estimate for the mean value $\langle \Delta m^{*} \rangle$ coincides with that of $\tilde m$ (see Supplemental Material). Thus, the CDF becomes narrowly peaked for large $T$ and $N$ around $\Delta m^{*} \sim (2 \mu T \ln N)^{1/2}$.

\begin{figure}
\includegraphics[width=0.48\columnwidth]{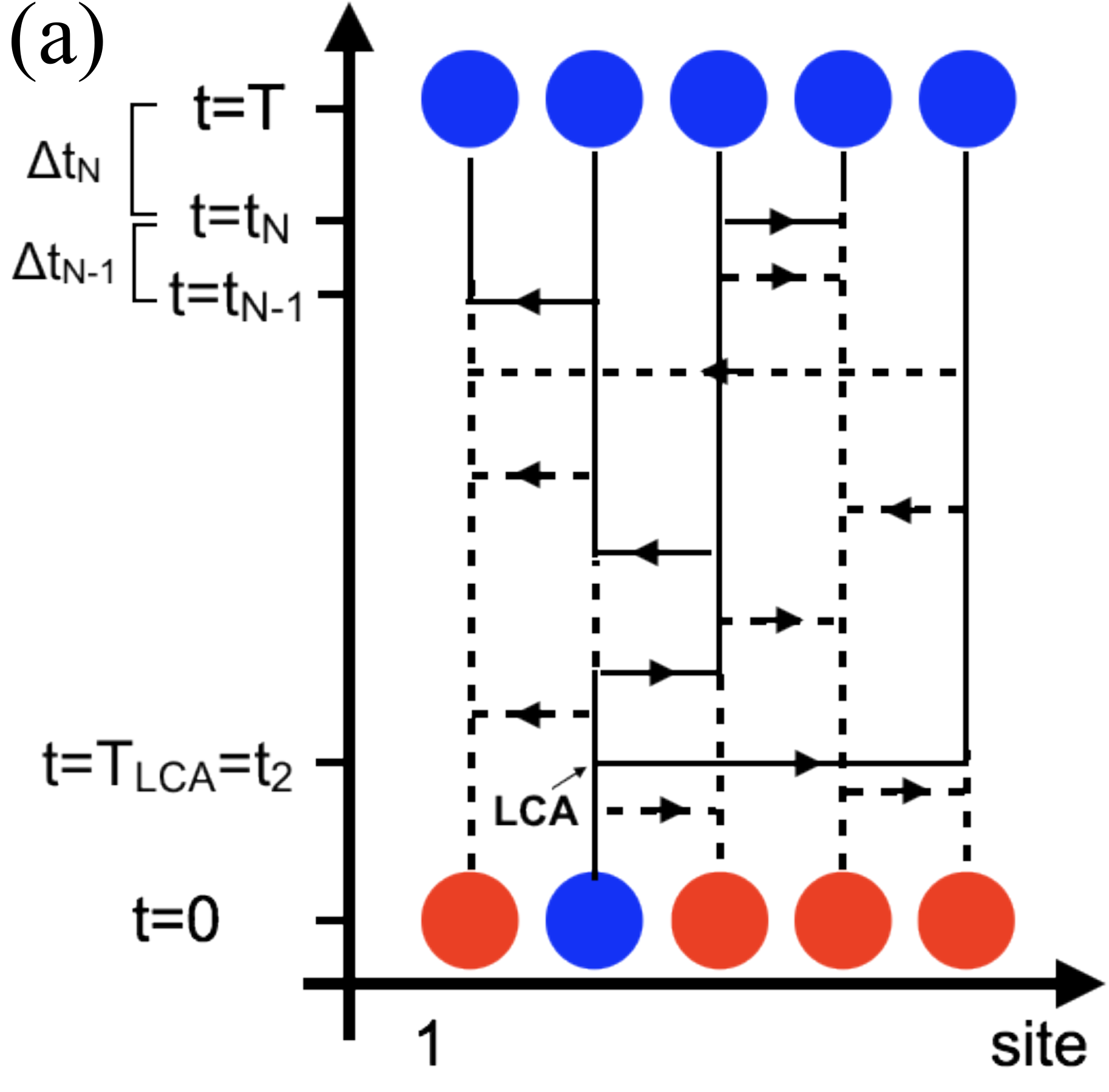}
\includegraphics[width=0.48\columnwidth]{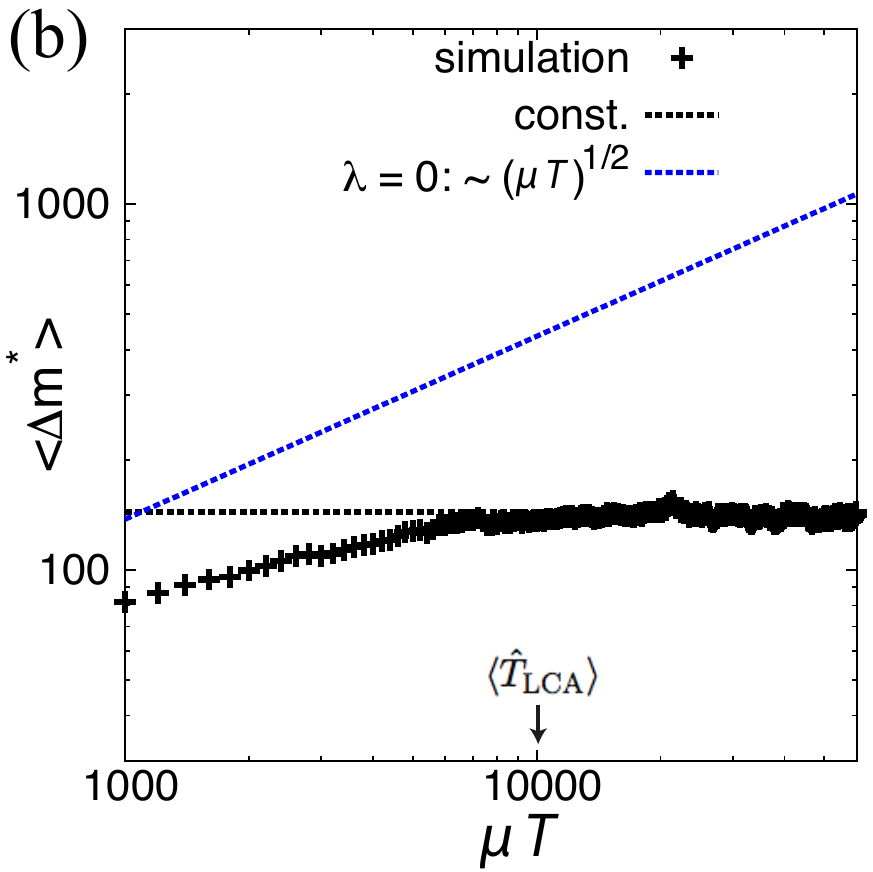}
\caption{\label{genealogy_fig} The genealogy and its implications. (a) Illustration of the history of mutation accumulation. Vertical lines represent mutational paths. Horizontal arrows mark loss/replacement events by which one cell copies its mutational configuration, in the direction of the arrow. Dashed lines are mutational paths that are lost, while bold lines are paths which survive until time $t=T$, constituting the genealogy. If $T$ is large enough, the genealogy possesses a single root, the last common ancestor (LCA). (b) Mean difference between maximum mutation number  and the population mean, $\langle \Delta m^* \rangle = \langle m^{*} \rangle - \langle m_{i} \rangle$, as a function of $\mu T$, for fixed $N=10,000$, with $\lambda=\mu$. Points are results from Monte Carlo simulations; the dashed black line illustrates the saturation constant. The blue dashed line marks the scaling prediction for $\lambda=0$, $\langle \Delta m^* \rangle \sim (\mu T)^{1/2}$.}
\end{figure}

In the case of a non-zero cell loss/replacement rate, $\lambda>0$, any two cells may have a common ancestor and thus do not accumulate mutations independently. It is then instructive to consider the \emph{genealogy} of the cell population, as illustrated in Fig. \ref{genealogy_fig}a. The genealogy describes the mutational history of all ancestors of cells at time $t=T$ and has the form of a binary tree, where branches connect daughter cells with their mothers \cite{Kingman1982a}. It contains all mutational paths that start at $t=0$ and reach the present. In considering the mutational statistics at time $t=T$, it is therefore sufficient to consider only mutations that occur on the genealogy \cite{Kingman1982a,genealogies_hudson}. 

The tree structure of the genealogy is characterized by its branching times $t_k$, at which the branch number changes from $k-1$ to $k$ (see Fig. \ref{genealogy_fig}a), i.e. during the period $t_k < t < t_{k+1}$, the genealogy consists of $k$ branches. The branching times can be determined by following the genealogy backwards in time $\hat t =T-t$; a coalescent process \cite{Kingman1982,genealogies_hudson,coalescent_theory_book}. This results in branching times whose intervals $\Delta t_k := t_{k+1}-t_k$ are exponentially distributed, with ${\rm Prob}(\Delta t_k) = \langle \Delta t_k \rangle^{-1} e^{-\Delta t_k/\langle \Delta t_k \rangle}$ and mean branching times (see Supplemental Material)
\begin{align}
\label{bar_t_k}
\langle \Delta t_k \rangle= \frac{N}{k \, (k-1)} \frac{1}{\lambda}\,\,\,\, .
\end{align}
Importantly, the accumulation of mutations along a single branch follows a simple, independent Poisson process \cite{genealogies_hudson} and thus the mean mutation number is simply $\langle m_i \rangle(T) =\mu T$. 


For times $T$ large enough, if one follows the coalescent process far backwards in time $\hat t$, the number of branches reduces until $\hat t=\hat T_{LCA} :=T-t_{2}$, beyond which there is only a single one left, the \emph{last common ancestor (LCA)} (see Fig. \ref{genealogy_fig}a). Thus, an LCA exists whenever $T>\hat T_{\rm LCA}$, which is on average $\langle \hat T_{\rm LCA} \rangle = \sum_{k=2}^N \langle \Delta t_k \rangle \approx N/\lambda$ (for $N \gg 1$).
In that case, before the time $T_{\rm LCA} = T - \hat T_{\rm LCA} = t_{2}$, the genealogy corresponds to the mutational path of a single cell for which the maximum $m^*$ equals the mutation number $m$. Hence, it follows that $\Delta m^{*} = m^{*} - \langle m_{i} \rangle >0$ only for times larger than $T_{LCA}$, such that the statistics of $\Delta m^{*}$ does not explicitly depend on the total time $T$ if $T > \hat T_{\rm LCA}$. Indeed, Monte Carlo simulations of the model confirm this conjecture for $\langle \Delta m^{*} \rangle(T)$, as is illustrated in Fig. \ref{genealogy_fig}b for a high mutation rate $\mu=\lambda$, where a plateau is reached around $T \sim \langle \hat T_{\rm LCA} \rangle$. Note that this is in contrast to the case of purely asymmetric divisions, $\lambda=0$, for which $\langle \Delta m^* \rangle \sim (\mu T)^{1/2}$. 

To support this finding for $T>\hat T_{\rm LCA}$ quantitatively, we note that the branching times are random and exponentially distributed. Thus, the branching of the genealogy corresponds to a Markov process, a branching process with initially two branches at time $T_{\rm LCA}$ with branching rate \emph{per branch} $\nu_k : = 1/\langle \Delta t_k \rangle k$. By approximating the random accumulation of mutations along each branch (Poisson process) by diffusive random walks in the variables $m_{i} - \mu T$ (valid for $\mu T \gg 1$), the mutation accumulation of the genealogy becomes an unbiased \emph{branching random walk (BRW)}. For unit branching rate, it has been shown \cite{branching_RW_max_bramson_1978} that the CDF of the maximum $\Delta m^{*}$ of the BRW, $P^{*}_{N}(\Delta m_{c},\tau)={\rm Prob}(\Delta m^{*}(\tau) \leq \Delta m_{c})$, follows a Fisher-KPP-type equation  \cite{Fisher1937,Kolmogorov1937}
\begin{align}
\label{Fisher-KPP_eq}
\partial_\tau P^*_{N} = D \, \frac{\partial^{2}P^*_{N}}{\partial \Delta m_{c}^{2}}  - P^*_{N}[1-P^*_{N}] \,\,\, ,
\end{align}
with the dimensionless time $\tau=\nu t$ measured in units of the constant branching time $\nu^{-1}$, and $D$, the diffusion constant of the random walk. The solution of this equation has the form
\begin{align}
\label{FKPP_sol}
P^*_{N}(\Delta m_{c},\tau) &= f(\Delta m_{c} - \tilde m(\tau)) 
\end{align}
with the median of $P^*_{N}$, $\tilde m(\tau) = 2\sqrt{D}\,\tau + O(\ln\tau)$ \cite{branching_RW_max_bramson_1978}.

Here, the branching rate $\nu_k$ is not constant. By a step-wise rescaling of time in units of branching times as  $\tau(t) := \nu_k (t-t_k)+ \tau_{k}$, with $\tau_k = \sum_{k'=2}^{k-1} \nu_{k'} \Delta t_{k'}$ for the largest $t_{k} < t$, $k>2$, the corresponding BRW in the time scale $\tau$ has unit branching rate and effective diffusion constant $D_{k} := \mu/2\nu_{k}$. While $D_k$ does not explicitly depend on time, we can take the ensemble average over the branch numbers $k$, $D(\tau) := \langle D_k \rangle_k|_\tau = \mu N/(2\lambda) \times \langle(k-1)^{-1} \rangle_k|_\tau$, to get an effective time-dependent diffusion constant. According to Ref. \cite{Fang2012} (see also Supplemental Material), for a diffusion constant $D(\tau')$ that decreases over time $\tau' < \tau = \tau(T)$, the CDF of the maximum of a BRW has the form of a Fisher-KPP wave, according to (\ref{FKPP_sol}), but with
\begin{align}
\label{tilde_m_eq}
\tilde m(\tau) &= \left[\int_0^\tau 2\sqrt{D(\tau')} \, d\tau' \right] (1 - O(\tau^{-\frac{2}{3}})) \approx \mathcal C_{\tilde m} \sqrt{\frac{\mu N}{\lambda}} \,\,\, ,
\end{align}
where $\mathcal C_{\tilde m}=\int_{0}^{\infty} \sqrt{2\langle (k-1)^{-1} \rangle_{k}|_{\tau'}} \, d\tau'$ is independent of the model parameters $N,T,\lambda$ and $\mu$. Here, we took the limit $\tau(T) \to \infty$, which is valid for large $N$, as this is a unit rate branching process with branch number $k(T) = N \approx e^{\tau}$. Thereby, terms of $O(\tau^{-\frac{2}{3}})$ are omitted, and the integral becomes independent of $N$. A numerical evaluation (see Supplemental Material) yields $\mathcal C_{\tilde m} \approx 1.79$.

\begin{figure}
\includegraphics[width=0.49\columnwidth]{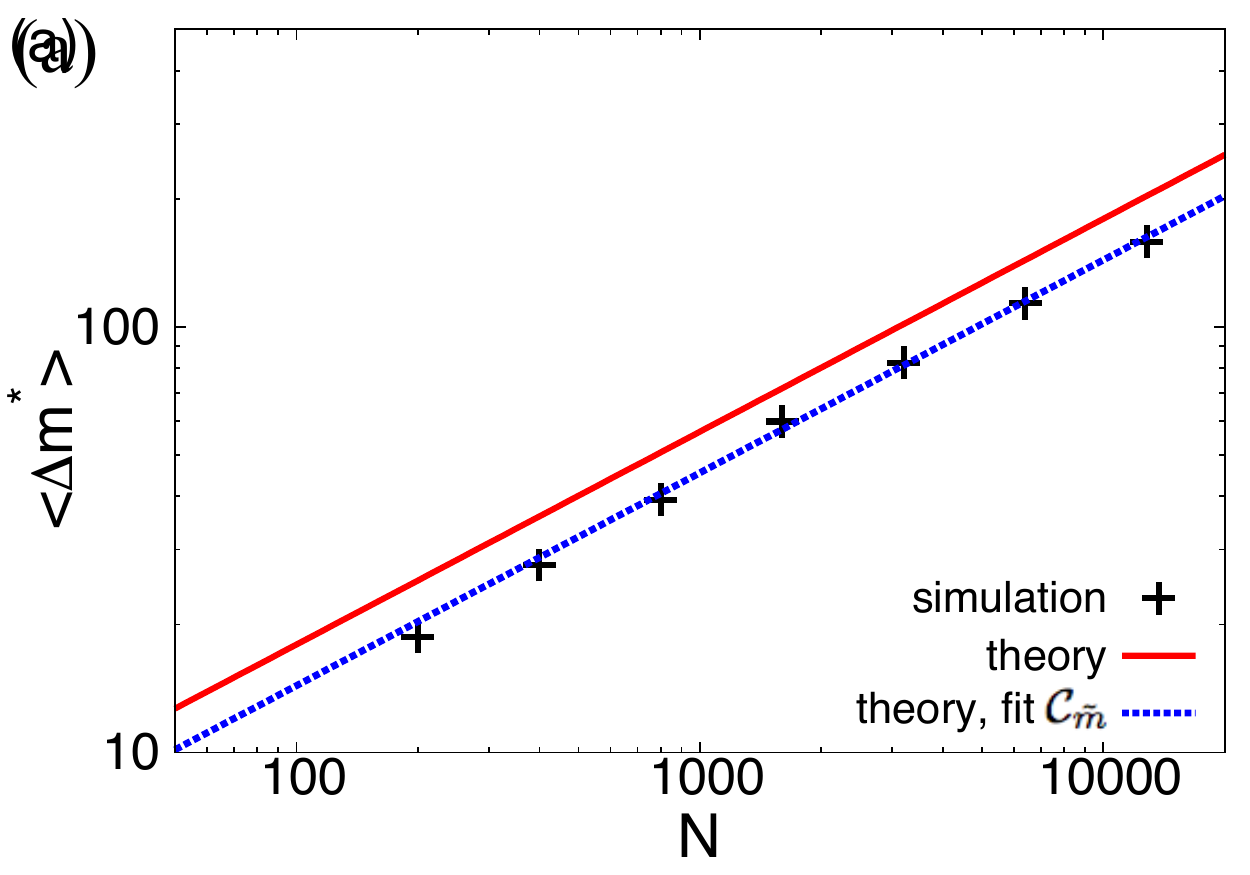}
\includegraphics[width=0.49\columnwidth]{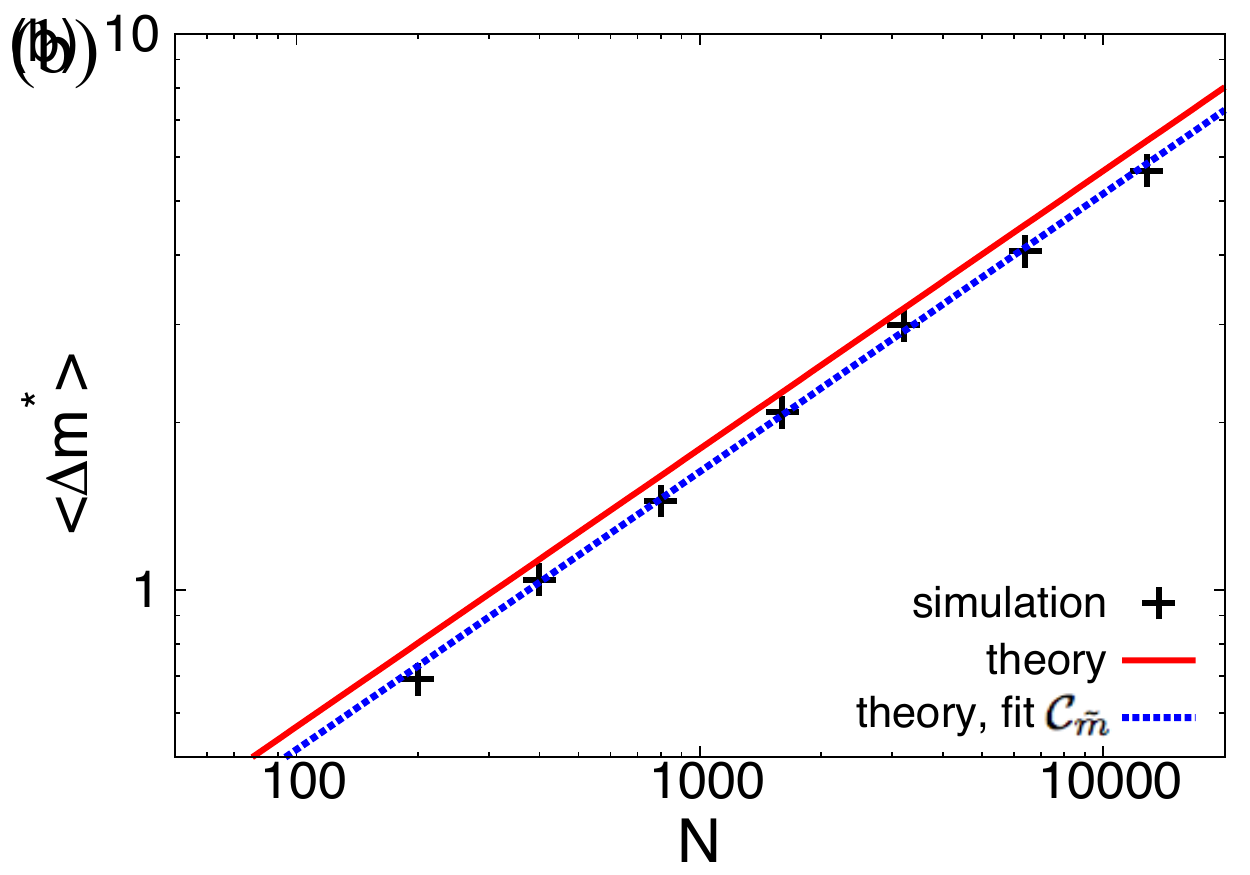}
\caption{\label{mean_max_N_T=10N_fig} Mean maximum mutation number ahead of the mean, $\langle \Delta m^{*} \rangle $ as a function of $N$, for $T=10 \, N/\lambda$ such that $T > \hat T_{\rm LCA}$. Shown are the results of Monte Carlo simulations (pluses), and theoretical predictions from the BRW approximation, Eq. (\ref{tilde_m_eq}), for fitted numerical constant $\mathcal C_{\tilde m}^{\rm fit}$ (blue dashed line) and theoretically estimated value (solid red line), $\mathcal C_{\tilde m}^{\rm th} =1.79$ for (a) $ \mu=\lambda$ ($\mathcal C_{\tilde m}^{\rm fit} = 1.43$) and (b) $\mu=0.001 \lambda$ ($\mathcal C_{\tilde m}^{\rm fit} = 1.63$) .}
\end{figure}

As expected, $\tilde m$ becomes independent of $T$ if an LCA exists. The mean, $\langle \Delta m^{*} \rangle$, follows the same scaling in $N$ and $T$, since due to the wave property of the CDF it only differs by a constant. This confirms our previous conjecture and the simulation results for $\langle \Delta m^{*} \rangle(T)$ (Fig. \ref{genealogy_fig}b). In Fig. \ref{mean_max_N_T=10N_fig}, we compare the theoretical results from Eq. (\ref{tilde_m_eq}) with results for $\langle \Delta m^{*} \rangle$ from Monte Carlo simulations, as a function of $N$. Here, $T$ was scaled with $N$ ($T=10 \, N/\lambda$) to assure that $T>\hat T_{\rm LCA}$. The theory with fitted numerical constant $\mathcal C_{\tilde m}$ ($\mathcal C_{\tilde m}^{\rm fit}$, blue dashed line) shows excellent agreement with simulations, while the calculated value ($\mathcal C_{\tilde m}^{\rm th} = 1.79$, red line) shows some deviation. These deviations are expected due to contributions with small $\tau'$ in $\mathcal C_{\tilde m}$ (whose approximation is valid for large $\tau'$). Remarkably, the predictions of the approximation are also valid for $\mu T \sim 1$ as shown in Fig. \ref{mean_max_N_T=10N_fig}b for $\mu=0.001\,\lambda$. 

While the nonlinear form of the Fisher-KPP equation does not admit an exact solution, the CDF's tail with $\bar P_{N} = 1-P^*_{N}(\Delta m_{c},\tau) \ll 1$ can be mapped onto a simple diffusion equation with time-varying diffusion constant (see Supplemental Material). Since variances add linearly in this case, for $T>\hat T_{LCA}$ and $\Delta m_{c} \gg \tilde m$, the CDFs tail is that of a non-normalized Gaussian function,
\begin{align}
\label{CDF_BRW_tail}
P^*_{N}(\Delta m_{c}) \sim 1- \frac{N \sigma_{\rm eff} \, e^{-\frac{\Delta m_{c}^2}{2 \sigma_{\rm eff}^2}}}{\sqrt{2\pi}\, \Delta m_{c}} \,\,\, ,
\end{align}
with
\begin{align}
\label{sigma_eff_eq}
\sigma_{\rm eff}(\tau) =2 \sqrt{ \int_0^\tau D(\tau') \,d\tau'} \approx \mathcal C_\sigma \, \sqrt{\frac{\mu N}{\lambda}} \,\,\, ,
\end{align}
where $\mathcal C_{\sigma} \approx \sqrt{\int_{0}^{\infty} 2 \langle (k-1) \rangle_{k}|_{\tau'} \, d\tau'} \approx 0.76$ (see Supplemental Material). 


%

For $T< \hat T_{\rm LCA}$ the population may not possess an LCA. In this case the genealogy fragments into $k$ sub-genealogies $l=1,...,k$, each representing a subpopulation of $n_l$ cells, while $N=\sum_{l=1}^k n_l$. Each sub-genealogy, however, can again be approximated by a branching random walk, with a CDF $P_{l}(\Delta m_{c})$ having a Gaussian tail according to Eq. (\ref{CDF_BRW_tail}). Therefore, and since the subpopulations accumulate mutations independently from each other, the CDF of the whole population, $P^{*}_{N}(\Delta m_{c})$, is approximated for large $N$ by a Gumbel distribution according to Eq. (\ref{gumbel_eq}) \cite{extr_val_bouchaud}, scaled by $\sigma_{\rm eff}$, and with effective number of independently distributed random variables, $k \approx N/\lambda T$ (see Supplemental Material). This CDF has then median and scaling width,
\begin{align}
\label{m^*_N_fixed-T}
\tilde m & \approx \mathcal C_{\sigma} \sqrt{2\, \ln\left(\frac{N}{\lambda T}\right)\mu T}, \hspace{2mm} \sigma_{N} \approx \mathcal C_{\sigma}\sqrt{\frac{ \mu T}{2 \ln\left(\frac{N}{\lambda T}\right)}} ,
\end{align}
where $\mathcal C_{\sigma}$ is according to Eq. (\ref{sigma_eff_eq}). The same scaling applies to $\langle \Delta m^{*} \rangle$.

\begin{figure}
\includegraphics[width=0.49\columnwidth]{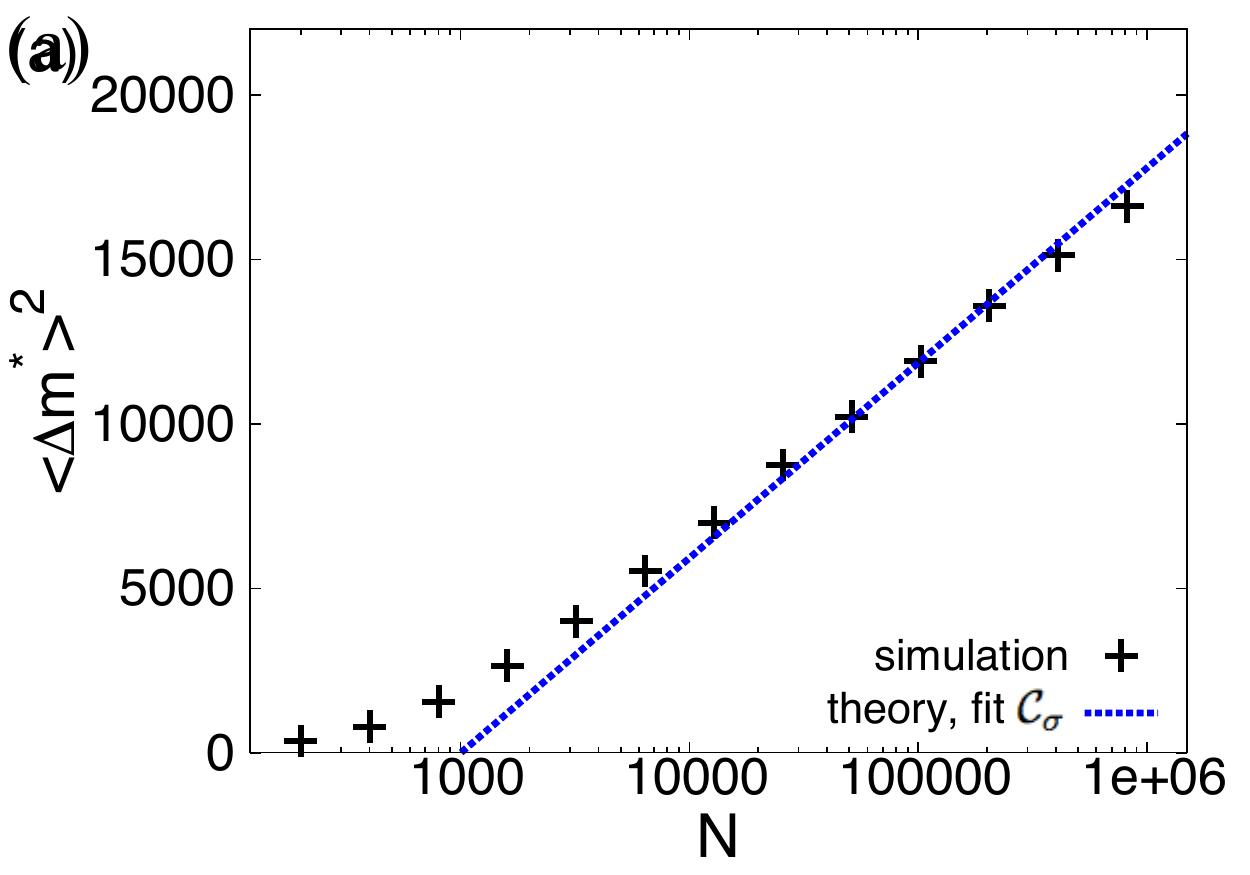}
\includegraphics[width=0.49\columnwidth]{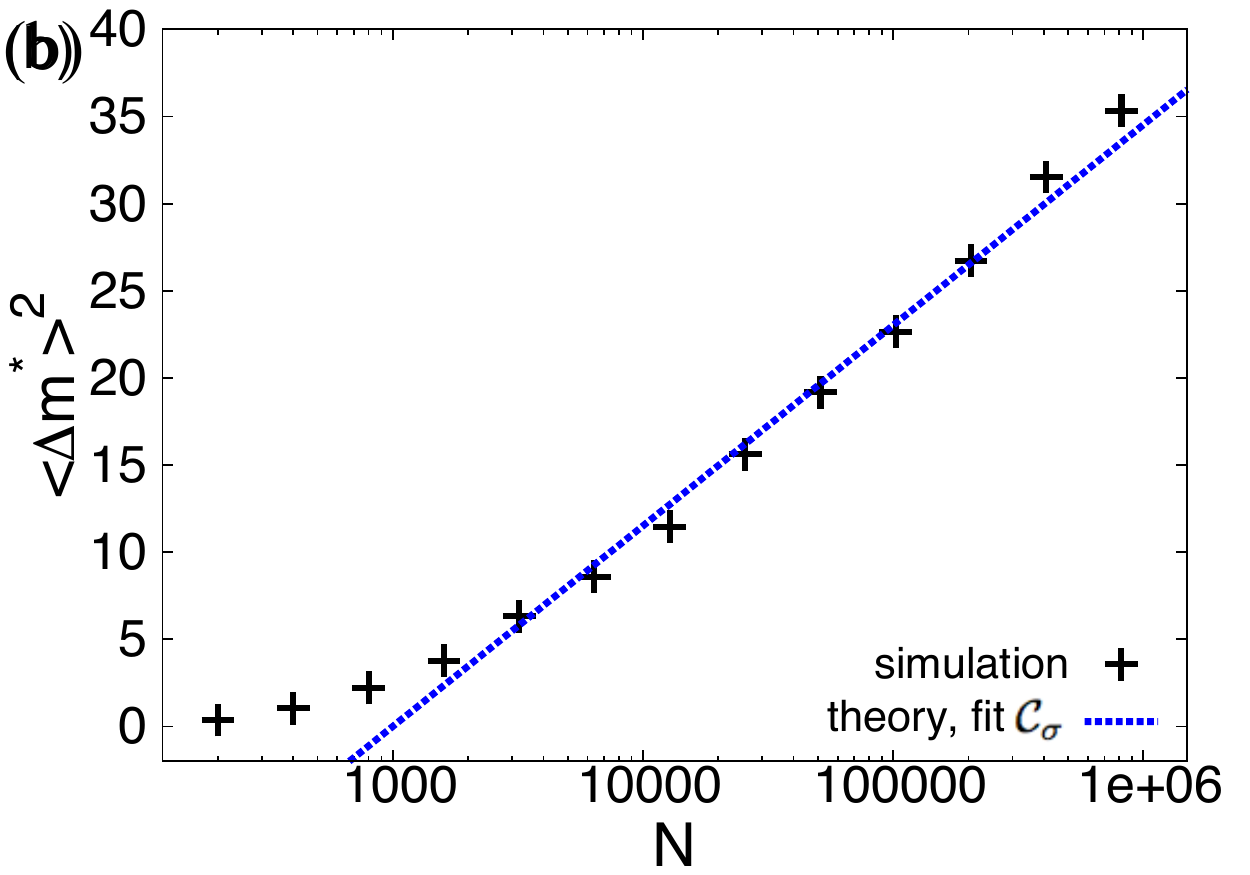}
\caption{\label{mean_max_N_T=1000_fig} Squared mean maximum mutation number ahead of the population mean, $\langle \Delta m^{*} \rangle^{2}$ as a function of $N$, for fixed $T=1000/\lambda$. Shown are the results of Monte Carlo simulations (pluses) and the theory, Eq. (\ref{m^*_N_fixed-T}), with fit parameter $\mathcal C_{\sigma}^{\rm fit}$ (dashed line) for (a) $\mu=\lambda$ ($\mathcal C_{\sigma}^{\rm fit} = 1.13$) and (b) $\mu=0.001\lambda$ ($\mathcal C_{\sigma}^{\rm fit} = 1.59$).}
\end{figure}

Thus, we expect the scaling $\Delta m^{*} \sim \sqrt{\ln{N}}$, as for $\lambda=0$, but with a negative offset $-2\mathcal C_{\sigma}^{2}\ln(\lambda T)\mu T$ under the square root. Figure \ref{mean_max_N_T=1000_fig} shows Monte Carlo simulations of $\langle \Delta m^{*} \rangle$ together with the theory, with fitted $\mathcal C_{\sigma}^{\rm fit}$, which shows a good agreement in the shown range of $N$, for both large and small mutation rates, $\mu=\lambda$ and $\mu=0.001\lambda$. Deviations from the theoretically approximated value $\mathcal C_{\sigma}^{\rm th}=0.76$ are expected, for the same reasons as for $\mathcal C_{\tilde m}$ before, and furthermore for very large $N$, since the extreme value distribution of Poisson variables differs from the Gaussian approximation in this limit.

\begin{figure}
\includegraphics[width=0.49\columnwidth]{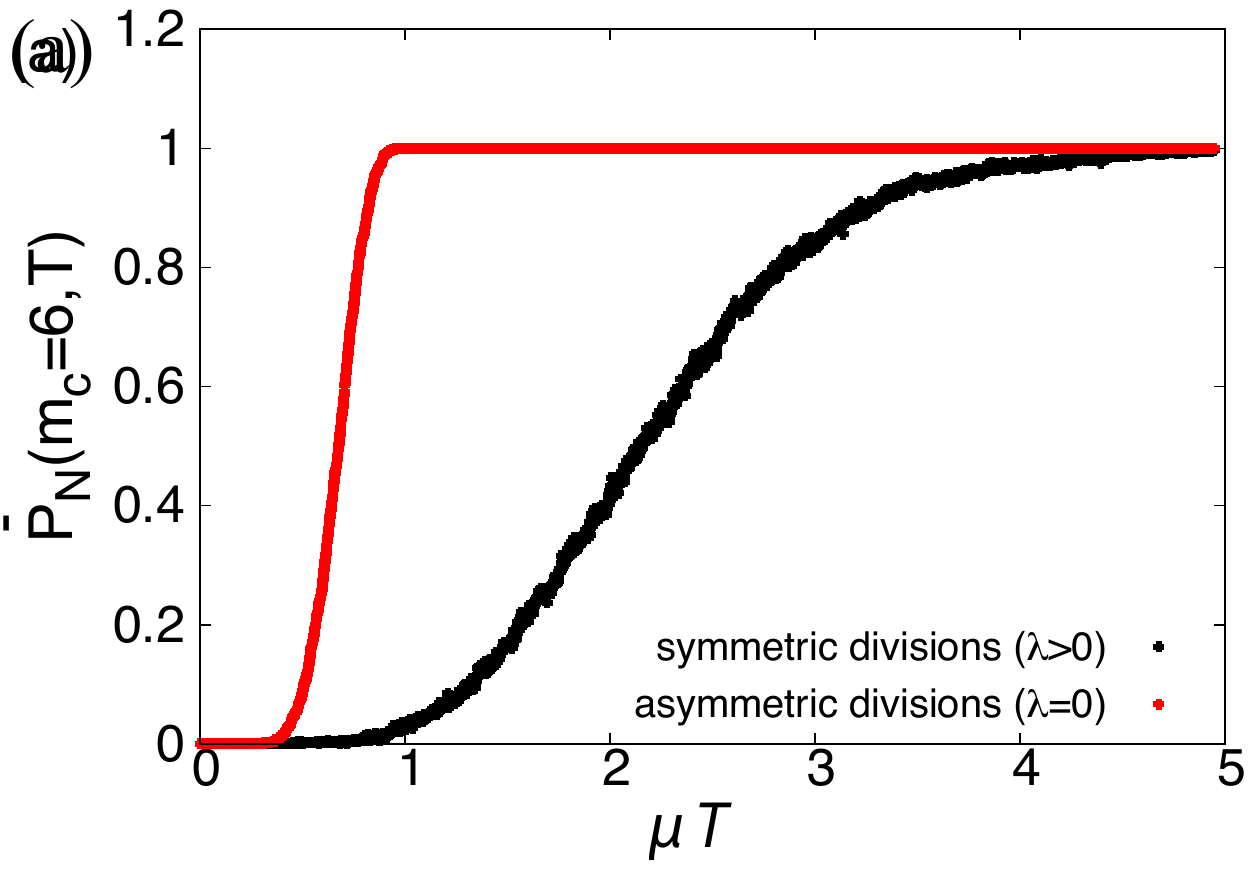}
\includegraphics[width=0.49\columnwidth]{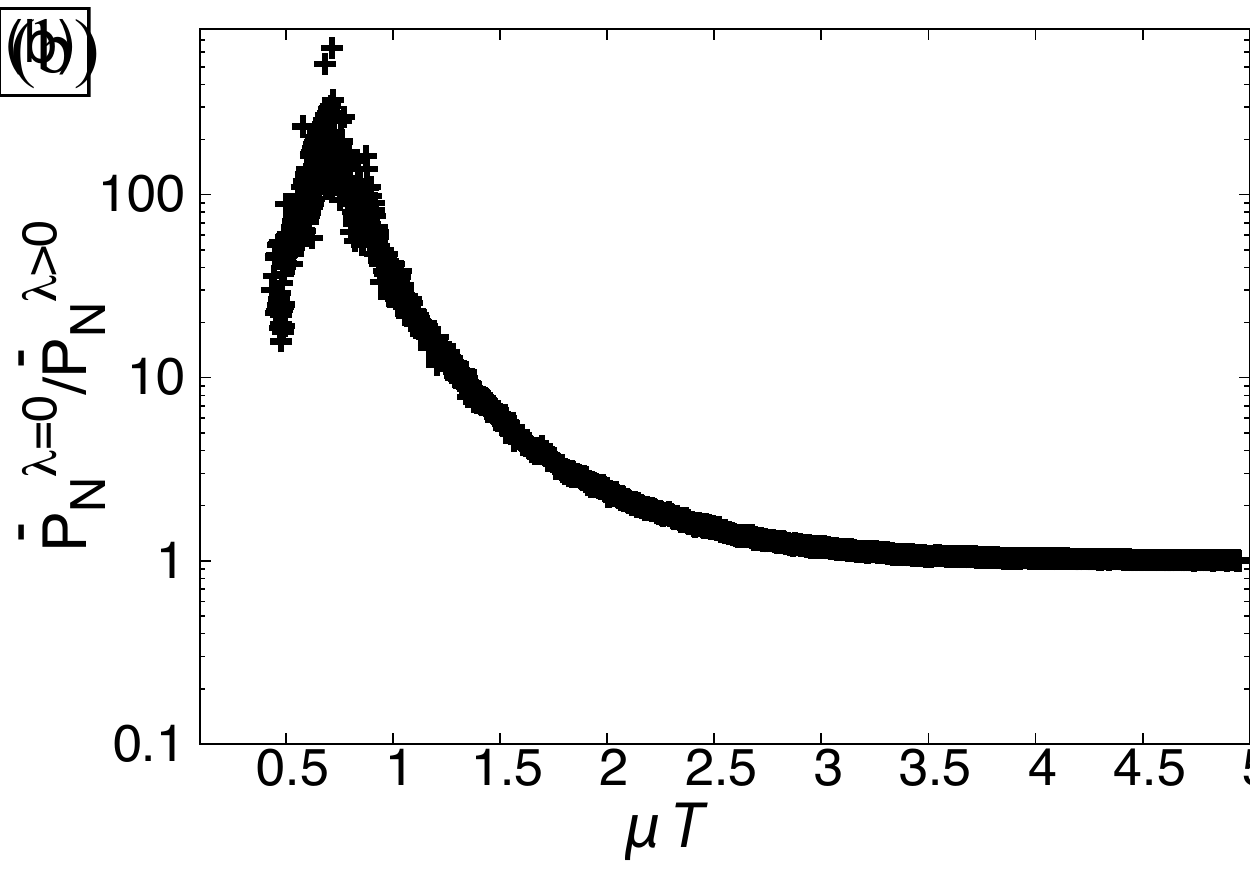}
\caption{\label{P_t_fig} Probability $\bar P_{N}(T)$ of accumulating a critical number of $m_{c}=6$ mutations in an area of 0.1 ${\rm cm}^{2}$, corresponding to $10^{5}$ cells. (a) $\bar P_{N}$ as function of time for physiological parameters of human eyelid \cite{martincorena_jones_science2016,simons_pnas_2016}, with $\mu = 0.27/(63\, \mbox{years})$ and $\lambda = 0.5/\mbox{week}$, corresponding to $\lambda \approx 6000\mu$ (black points), and for the hypothetical case $\lambda=0$ (red points). (b) The risk ratio $(\bar P_{N})^{\lambda=0}/(\bar P_{N})^{\lambda>0}$ for the two scenarios from (a).}
\end{figure}

Finally, we consider the risk of accumulating a critical mutation number $m_{c}$, $\bar P_{N}(m_{c},T) = 1 - P_{N}^{*}(m_{c},T)$. In Figure~\ref{P_t_fig}, results from stochastic simulations for $\bar P_{N}(T)$ are shown for $m_{c}=6$ \cite{armitage_doll_1954,mutrate_cancer_tomlinson_pnas1996} with parameters chosen to match physiological conditions of human epidermal stem cells \cite{martincorena_jones_science2016,simons_pnas_2016} (see figure caption), comparing a model of stochastic stem cell loss and replacement, $\lambda>0$, with the hypothetical case of asymmetric stem cell divisions, $\lambda=0$. As the front of $P^{*}_{N}(m_{c})$ moves with a speed $\mu + \partial_{t} \tilde m$, which is larger for $\lambda=0$ (see Figure \ref{genealogy_fig}b), it reaches the critical value $m_{c}=6$ earlier than for $\lambda>0$, resulting in an earlier increase of $\bar P_{N}(m_{c},T)$ in Figure \ref{P_t_fig}a. Remarkably, the threshold of $m_{c}=6$ mutations is already frequently exceeded for mean mutation numbers $\mu T<1$, which demonstrates that the acquisition of a critical number of mutations is indeed dominated by extreme values. In Figure \ref{P_t_fig}b, the \emph{risk ratio} between the risk $\bar P_{N}|_{\lambda=0}$ for asymmetric divisions and that for symmetric divisions, $\bar P_{N}|_{\lambda>0}$, is shown. One observes a non-monotonic behavior: for intermediate times the
ratio is large, while it approaches one for large times. The initial growth of the ratio is due to the smaller slope in the tail for $\lambda > 0$, $1/\sigma^{\lambda>0}_{N} \sim \sqrt{\ln(N/(\lambda T))}$ compared to the slope $1/\sigma^{\lambda=0}_{N} \sim \sqrt{\ln(N)}$ for $\lambda=0$. Once $\bar P_{N}|_{\lambda=0}$ has saturated, however, $\bar P_{N}|_{\lambda>0}$ catches up with the latter until both approach $\bar P_{N}=1$. This provides a theoretical foundation for claims that the risk of tumor initiation is decreased by stochastic stem cell loss and replacement \cite{Shahriyari2013,lander_mut-accumul}, yet this advantage is limited to intermediate time scales.

Until now, we have considered an ``infinite-dimensional'' process in which any cell may replace another. However, the majority of biological tissues are low-dimensional, where stem cell loss and replacement occur between neighboring cells in tubular (one-dimensional), epithelial (two-dimensional) or volumnar (three-dimensional) settings. Such situations can be modeled by embedding cells on a $d$-dimensional regular lattice, allowing replacement only between neighboring cells \cite{Klein2008,Klein2011}. In this case, our general theory remains valid, based on the mapping of the genealogy on a BRW; only the distribution of branching times, $\Delta t_{k}$, differs (see Supplemental Material). Nonetheless, only for $d=1$ and $T>\hat T_{\rm LCA}$, a significantly different scaling with $\tilde m \sim N \sqrt{\mu/\lambda}$ is observed, compared to the infinite-dimensional case.

In summary, we have studied the asymptotic behavior, with time $T$ and cell number $N$, of the maximum mutation number statistics in a renewing cell population, in which cells may be stochastically lost and replaced (Moran process). This is of importance if multiple neutral or near-neutral mutations can cooperate through epistasis to trigger hyper-proliferation, a potential tumor initiating event. We showed that, for a non-zero cell replacement rate, $\lambda>0$, the difference between the average maximum mutation number and the population mean, $\langle \Delta m^{*} \rangle$, saturates to a constant value when scaled with $T$. Using an analogy to branching random walks, we showed that the value of this constant scales as $(\mu N/\lambda)^{1/2}$ for $\lambda>0$, while in the absence of symmetric loss/replacement, $\lambda=0$ (e.g. for asymmetric stem cell divisions only), $\langle \Delta m^{*} \rangle$ scales with time and cell number as $(\mu T \ln(N))^{1/2}$. If $T$ is fixed, $\langle \Delta m^{*} \rangle$ scales as $(\mu T \ln(N/(\lambda T)))^{1/2}$ for $\lambda>0$ and large $N \gg \lambda T$. From this result, it follows that at intermediate time scales, the risk of triggering hyper-proliferation is higher for asymmetric than for symmetric stem cell divisions, yet these probabilities converge for large times. Crucially, our theory also applies in a low-dimensional setting, where cell loss and replacement occurs between neighbors, albeit with a different scaling dependence for one-dimensional structures.


We thank Steffen Rulands for valuable discussions and for help with literature research on genealogies and the coalescent process. We further are thankful for the support by a EPSRC Critical Mass Grant and a DFG Research Fellowship.



\end{document}